\documentclass[twocolumn,showpacs,preprintnumbers,prd]{revtex4-1}
\usepackage{hyperref}
\usepackage{rotating}
\usepackage{amsmath,graphicx,amssymb}
\usepackage{bm}

\newcommand\Br{{\rm Br}}
\renewcommand\l{\left(}
\renewcommand\r{\right)}

\newcommand{\be}{\begin{equation}}
\newcommand{\ee}{\end{equation}}
\newcommand{\bear}{\begin{eqnarray}}
\newcommand{\ear}{\end{eqnarray}}
\newcommand{\ba}{\begin{array}}
\newcommand{\ea}{\end{array}}

\begin{document}

\author{S.V.~Demidov}\email{demidov@ms2.inr.ac.ru}
\author{D.S.~Gorbunov}\email{gorby@ms2.inr.ac.ru}

\affiliation{Institute for Nuclear Research of the Russian Academy of Sciences, 60th October Anniversary prospect 7a, Moscow 117312, Russia}


\title{
Flavor violating processes with sgoldstino pair production}

\date{\today}



\begin{abstract}

In supersymmetric extensions of the Standard Model of particle physics 
(SM), goldstino superpartners --- scalar and pseudoscalar sgoldstinos
--- can be light enough for emerging in decays of SM 
particles. Sgoldstino interaction with SM fields is suppressed by
the scale of supersymmetry breaking in the whole theory. Hence,
searches for sgoldstinos give an opportunity to probe the underlying
mechanism of supersymmetry breaking. Sgoldstino couplings to SM fields
are proportional to the supersymmetry breaking parameters --- MSSM
soft terms --- and therefore can lead to flavor violating processes in
quark and lepton sectors. We consider flavor violating processes
involving sgoldstino pair production which are driven by sgoldstino
couplings proportional to squark and slepton soft mass terms, $\tilde
m^2_{LL}$ and $\tilde m^2_{RR}$. We find that present limits on
off-diagonal entries in squark and slepton squared mass matrices allow
$t$-, $b$-, $c$-quark and $\tau$-lepton decays at levels available for
study with existing data (BaBar, Belle, CLEOc) and in ongoing
experiments (LHCb, CMS, ATLAS). In particular, we obtain the
following branching ratios $\Br\l t\to c SP\r\lesssim 10^{-7}$, $\Br \l \tau
\to \mu S P\r \lesssim 10^{-7}$, $\Br \l B_s\to S P \r \lesssim
10^{-4}$, $\Br \l 
B\to K^{(*)} S P \r \lesssim 10^{-4}$, $\Br \l D\to S P \r \lesssim 10^{-7}$ with
sgoldstino subsequent decays into kinematically allowed pairs of SM
particles $\gamma\gamma$, $e^+e^-$, $\mu^+\mu^-$, etc. Remarkably, the
prominent signature of sgoldstino pair production is two muon pairs
with pair momenta peaked at sgoldstino masses.
 
\end{abstract}
\maketitle


\section{Introduction}

Low energy supersymmetry provides a technically natural solution to
the gauge hierarchy problem and is still very attractive in spite of
absence of a clear experimental evidence for
superpartners. Supersymmetry if it exists, must be spontaneously
broken and hence there is a special Goldstone supermultiplet,
which includes scalar $S$ and pseudoscalar $P$ sgoldstinos, goldstino
and auxiliary bosonic field whose vacuum expectation value $F$ breaks
supersymmetry in the whole theory. Sgoldstinos being massless at
tree-level, get masses from higher order corrections. They can be
light if supersymmetry breaking happens at relatively low energy, as
in models with gauge mediation \cite{Giudice:1998bp} and in no-scale
supergravity \cite{Ellis:1984kd}.  

Sgoldstino phenomenology was extensively studied in literature (for a
brief review see~\cite{Gorbunov:2001pd}) and has got special attention
after HyperCP anomaly: observation of three muon pairs with invariant
mass $214.3\pm 0.5$\,MeV in hyperon decay $\Sigma\to p
\mu^+\mu^-$\,\cite{hep-ex/0501014}.  The anomaly may be explained as
pseudoscalar sgoldstino production in $\Sigma\to pP$ with subsequent
decay $P\to\mu^+\mu^-$\,\cite{hep-ex/0501014,Gorbunov:2005nu} (scalar 
sgoldstino should be somewhat heavier in this case, $m_S\gtrsim
300$\,MeV, to avoid limits from $K\to\pi\mu^+\mu^-$). Sgoldstino
  mass scale lower than the SM superpartner mass scale implies that
  selfinteraction in goldstino sector is somewhat weaker as compared
  to coupling responsible for mediation of supersymmetry breaking
  to the SM sector.
Sgoldstino
explanation can be tested in kaon \cite{Gorbunov:2005nu} and $D$- and
$B$-meson decays \cite{Demidov:2006pt} and recent searches with
negative  results \cite{Hyun:2010an,Abouzaid:2011mi} close some part
of the model parameter space.  

In this {\em Letter} we proceed further with light sgoldstino
phenomenology concentrating on flavor violating processes with
sgoldstino pair production. 

Indeed, there are two types of low energy interactions between
sgoldstino and matter (quark and lepton) fields. The type-I involves
single sgoldstino field, the type-II 
utilizes both scalar and pseudoscalar sgoldstinos. The type-I
couplings are proportional to so-called left-right soft terms in
squark (slepton) squared mass matrices, which are determined by
trilinear soft supersymmetry breaking terms. The type-II couplings are
proportional to left-left and right-right terms of the same
matrices. The left-right terms 
are naturally suppressed by fermion (quark and lepton) masses with
respect to other terms. However, the
type-II couplings are suppressed by additional factor $1/F$ with
respect to type-I couplings. For general low
  energy supersymmetry breaking models type-II couplings might
  dominate over type-I provided that soft trilinear parameters are
  additionally suppressed.
Note also, that
flavor violating patterns in the two
types of sgoldstino couplings are generally different. 

Type-I couplings are responsible for sgoldstino decays into light SM 
particles and give rise to single sgoldstino production. Possible
variants of type-I flavor-violating couplings and corresponding
phenomenology have been studied in 
detail\,\cite{hep-ph/0007325,Gorbunov:2000ht,Gorbunov:2000cz,Demidov:2006pt}. 
Type-II flavor-violating terms in sgoldstino lagrangian were addressed
only in Ref.\,\cite{hep-ph/0007325}, where annihilation of neutral
mesons $\pi^0$, $K_{L(S)}$, $D^0$, $B^0$ into sgoldstino pair have
been studies. In this {\em Letter} we extend the study of type-II
terms to the three-body decays of heavy mesons, $t$-quark and
$\tau$-lepton. 


\section{Lagrangian}
\label{Sec:Lagrangian}

The relevant for present study interaction terms of sgoldstino 
with SM fields (up- and down-type quarks $f_{U_i}$,
$f_{D_i}$ and charged leptons $f_{L_i}$, $i=1,2,3$) read\,\cite{hep-ph/0007325} 
\begin{widetext}
\begin{equation}
\label{flavor-violation}
\begin{split}
{\cal L} = \frac{1}{4F^2}(S\partial_{\mu}P - P\partial_{\mu}S)
\left((\tilde{m}^{LL^{2}}_{L_{ij}} \!+\! \tilde{m}^{RR^{2}}_{L_{ij}})
\bar{f}_{L_{i}}\gamma^{\mu}\gamma^5f_{L_{j}} +
(\tilde{m}^{LL^{2}}_{L_{ij}} \!-\! \tilde{m}^{RR^{2}}_{L_{ij}})
\bar{f}_{L_{i}}\gamma^{\mu}f_{L_{j}}
+ 
(\tilde{m}^{LL^{2}}_{D_{ij}} \!+\! \tilde{m}^{RR^{2}}_{D_{ij}})
\bar{f}_{D_{i}}\gamma^{\mu}\gamma^5f_{D_{j}} 
\right.
\\
\left.
+
(\tilde{m}^{LL^{2}}_{D_{ij}} \!-\! \tilde{m}^{RR^{2}}_{D_{ij}})
\bar{f}_{D_{i}}\gamma^{\mu}f_{D_{j}}
+ 
(\tilde{m}^{LL^{2}}_{U_{ij}} \!+\! \tilde{m}^{RR^{2}}_{U_{ij}})
\bar{f}_{U_{i}}\gamma^{\mu}\gamma^5f_{U_{j}} +(\tilde{m}^{LL^{2}}_{U_{ij}} \!-\! \tilde{m}^{RR^{2}}_{U_{ij}})
\bar{f}_{U_{i}}\gamma^{\mu}f_{U_{j}}\right)\;,
\end{split}
\end{equation}
\end{widetext}
where sum goes over $i,j=1,2,3$. 
Supersymmetry violating vacuum expectation value $F$ of auxiliary
component of goldstino supermultiplet 
has the dimension of squared mass and the scale
of supersymmetry breaking in the whole theory is of order
$\sqrt{F}$. Parameters $\tilde{m}_{(D,U,L)_{ij}}^{XX^2}$, $X=L,R$,
 are MSSM soft supersymmetry breaking terms entering 
squark and slepton squared mass matrices (e.g.,  
$\tilde{m}_{D_{11}}^{LL^2}$ is the squared soft mass term for
superpartner of left $d$-quark, etc).   

There are 
bounds on off-diagonal elements of matrices 
$\tilde{m}_{(D,U,L)}^{LL^2}$ and $\tilde{m}_{(D,U,L)}^{RR^2}$ 
coming from absence 
of FCNC processes and non-SM contributions to 
neutral pseudoscalar meson mixings. 
For numerical estimates of the
decay rates we use values of soft supersymmetry breaking terms 
which are in agreement  
with present constraints found in literature 
\cite{Jager:2008fc,arXiv:0907.5386,arXiv:0801.1833,arXiv:0910.2663}. 
It is convenient to define the dimensionless 
variables which parametrize the relative strength of flavor violation
as   
\be
\delta^{(D,U,L)}_{XX_{ij}} = \frac{\tilde{m}^{XX^{2}}_{(D,U,L)_{ij}}}{\tilde{m}^2}\,,
\ee
where $X=L,R$ and $\tilde{m}$ refers to a common mass scale of
superpartners. 

For numerical estimates we have to choose certain values of
supersymmetric parameters. 
Without any confirmed evidence for the presence of supersymmetry in
Nature we are free to choose them at will provided no
contradictions with experiment. 
Let us fix general scale of SUSY particle masses at 
$\tilde{m}=1000\,\mathrm{GeV}$,  
and put all $RR$-components of flavor violating
terms to zero, 
$\tilde{m}^{RR^2}_{(D,U,L)_{ij}}=0\;.$
All others are taken as follows
\begin{eqnarray}
\label{reference-D}
\delta^{D}_{{LL}_{13}} = 0.14,\;\;\;
\delta^{D}_{{LL}_{23}} = 0.2,\\ 
\label{reference-U}
\delta^{U}_{{LL}_{12}} = 0.06,\;\;\; 
\delta^{U}_{{LL}_{23}} = 0.3,\;\;\; %
\delta^{U}_{{LL}_{13}} = 0.3,\\ %
\label{reference-L}
\delta^{L}_{{LL}_{13}} = 0.1,\;\; 
\delta^{L}_{{LL}_{23}} = 0.1  
\end{eqnarray}
They are at upper bounds of phenomenologically allowed ranges;  
we have not found in literature relevant limits on
$\delta^{U}_{{LL}_{23}}$ and $\delta^{U}_{{LL}_{13}}$. Hereafter we
treat all the
values~\eqref{reference-D}-\eqref{reference-L}
as reference numbers only. 
To simplify the presentation further, below we plot decay branching
ratios for the special choice of supersymmetry parameters 
\begin{equation}
\label{reference-unnitarity}
\tilde{m}^{2}=F\;,
\end{equation}
which we call the unitarity limit.\footnote{Indeed, by definition,
  all soft supersymmetry breaking terms are proportional to $F$, since
they appear via transfer of supersymmetry breaking from the hidden
sector to SM and hence should not exceed $F$.} We comment on the
behavior of decay rates with change of model parameters in
Sec.\,\ref{Sec:Summary}.   

Sgoldstinos are even with respect to $R$-parity and hence
unstable. Apart of \eqref{flavor-violation}, there are other
interaction terms (see 
e.g.\,\cite{Gorbunov:2001pd,hep-ph/0007325}) 
responsible for the instability. 
Sgoldstinos decay into SM
particles and two-body decay mode dominates, see \cite{hep-ph/0007325}
for details. Generally, sgoldstino decays into $\gamma\gamma$,
$e^+e^-$, $\mu^+\mu^-$, $\pi^+\pi^-$ and other hadronic modes, if
kinematically allowed. Which mode dominates depends strongly on the
model parameters, but heavy leptons in final state are more natural,
than light ones (for discussion see
\cite{hep-ph/0007325,Gorbunov:2005nu}). Thus, for sgoldstino pair 
production processes one can expect two muon pairs as a very promising
signature to look for. Other interesting signatures include photon
pair(s) and light mesons.

\section{Decay ${\cal P} \to SP$}
\label{Sec:2-body}

\begin{figure}[!htb]
\begin{center}
\begin{tabular}{ll}
\includegraphics[width=0.5\columnwidth]{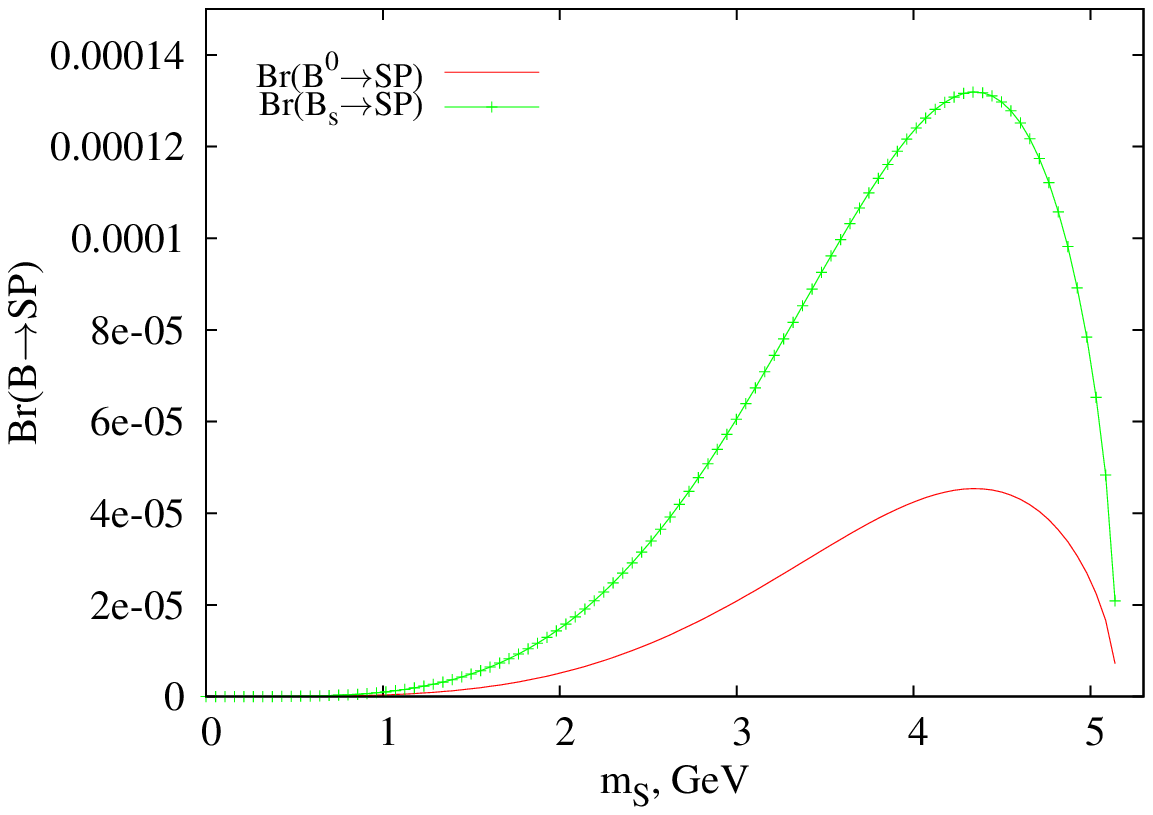}
&
\includegraphics[width=0.5\columnwidth]{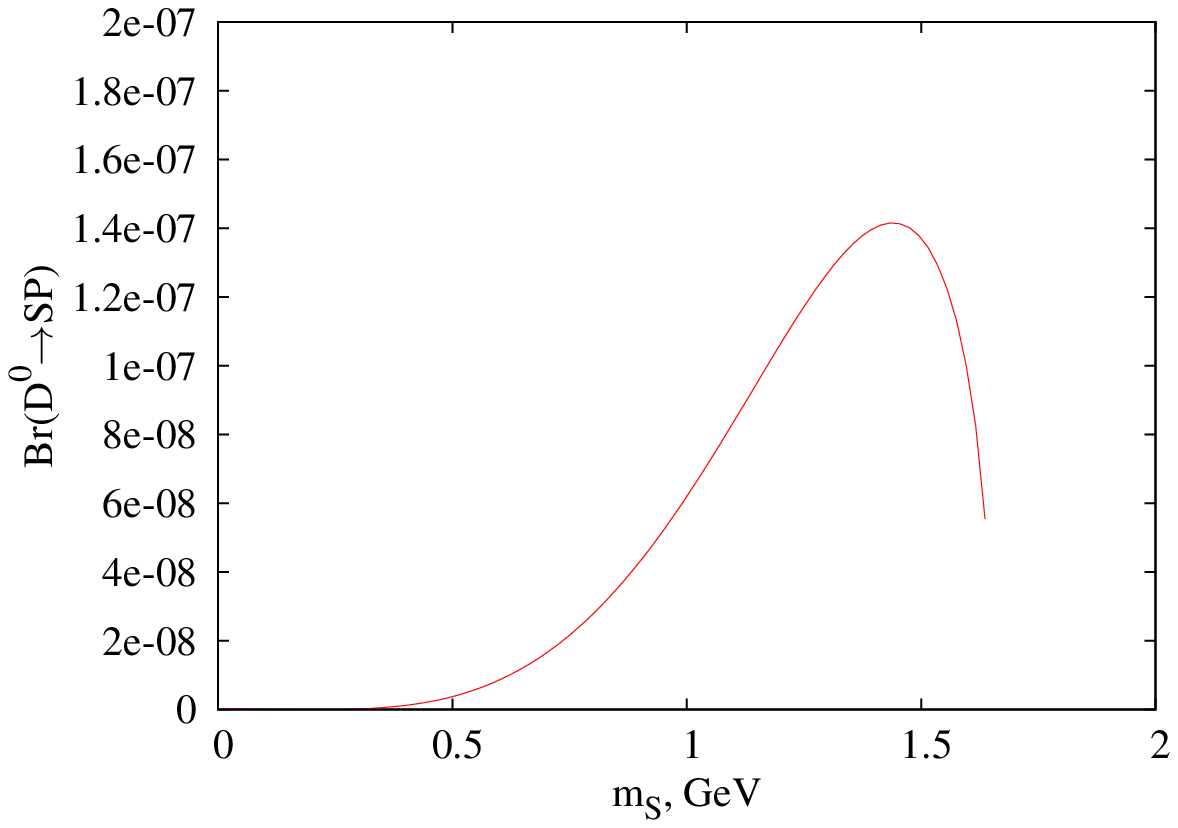}
\end{tabular}
\caption{\label{SP} Branching ratios of decays $B^0\to SP$ and
  $B_s\to SP$ (left) and $D^0\to SP$ (right) as functions of scalar
  sgoldstino mass $m_S$ for pseudoscalar sgoldstino mass fixed at
  $m_{P}=214$~MeV and other parameters fixed as explained in
  Sec.\,\ref{Sec:Lagrangian}.} 
\end{center}
\end{figure}

Here we consider two body decays of heavy neutral mesons into sgoldstino pair, 
$B^0\to SP$, $B_s\to SP$ and $D^0\to SP$. 
Decay rate of heavy pseudoscalar meson ${\cal P}$ with quark content
$\bar q_i q_j$ (i,j run over quark flavors) 
into sgoldstino pair due to 
interaction \eqref{flavor-violation} is given by 
the formula (c.f. Eq.(20) in \cite{hep-ph/0007325})
\begin{equation}
\label{1}
\begin{split}
\Gamma({\cal P}\to SP) &= \frac{f_{\cal P}^{2}}{M_{\cal P}}
\frac{\left|\tilde{m}_{(D,U)_{ij}}^{LL^{2}} +
  \tilde{m}_{(D,U)_{ij}}^{RR^{2}}\right|^2}{256\pi F^2} \\\times
\frac{(m_{S}^{2}-m_{P}^{2})^2}{F^2}
&\sqrt{\l 1+\frac{m_{P}^2-m_{S}^2}{M^2_{\cal P}}\r^2
- 4\frac{m_{P}^2}{M_{\cal P}^2}}\;,
\end{split}
\end{equation}
where  $M_{\cal P}$ and $f_{\cal P}$ are mass and decay constant of
pseudoscalar meson ${\cal P}$ (see Appendix
\ref{appendix-form-factors} for the numerical values adopted).  
For decay of antimeson $\bar {\cal P}$ (heavy meson with quark content 
$\bar q_j q_i$)  the same formula is valid
with substitution $i\leftrightarrow j$. For numerical estimates we 
adopt the set of parameter values given in Sec.\,\ref{Sec:Lagrangian}. 
The dependence of branching ratios of these decays on scalar sgoldstino
mass $m_S$ is depicted in Fig.~\ref{SP},    
where we put pseudoscalar sgoldstino mass to 
$m_{P}=214$\,MeV, as suggested by the sgoldstino explanation of 
HyperCP anomaly \cite{hep-ex/0501014,Gorbunov:2005nu}.  

Remarkably, the signal grows with sgoldstino mass, as the
corresponding matrix element exhibits the same behavior,
cf.~\eqref{flavor-violation},\eqref{1}. At large mass the branching
ratio decreases because of limited phase space. 

\section{Decays ${\cal P}\to \tilde{\cal P}SP$ and ${\cal P}\to {\cal V}SP$}
\label{Sec:3-body}

The same off-diagonal entries of squark squared mass matrix 
which determine the heavy meson  
two-body decays 
into sgoldstino pairs described in Sec.\,\ref{Sec:2-body} 
lead to three-body decays
of the same meson ${\cal P}$ to light pseudoscalar meson 
$\tilde{\cal P}$ and 
sgoldstino pair and to light vector meson 
$\tilde{\cal V}$ and 
sgoldstino pair. 

\begin{figure}[!htb]
\begin{center}
\begin{tabular}{ll}
\includegraphics[width=0.5\columnwidth]{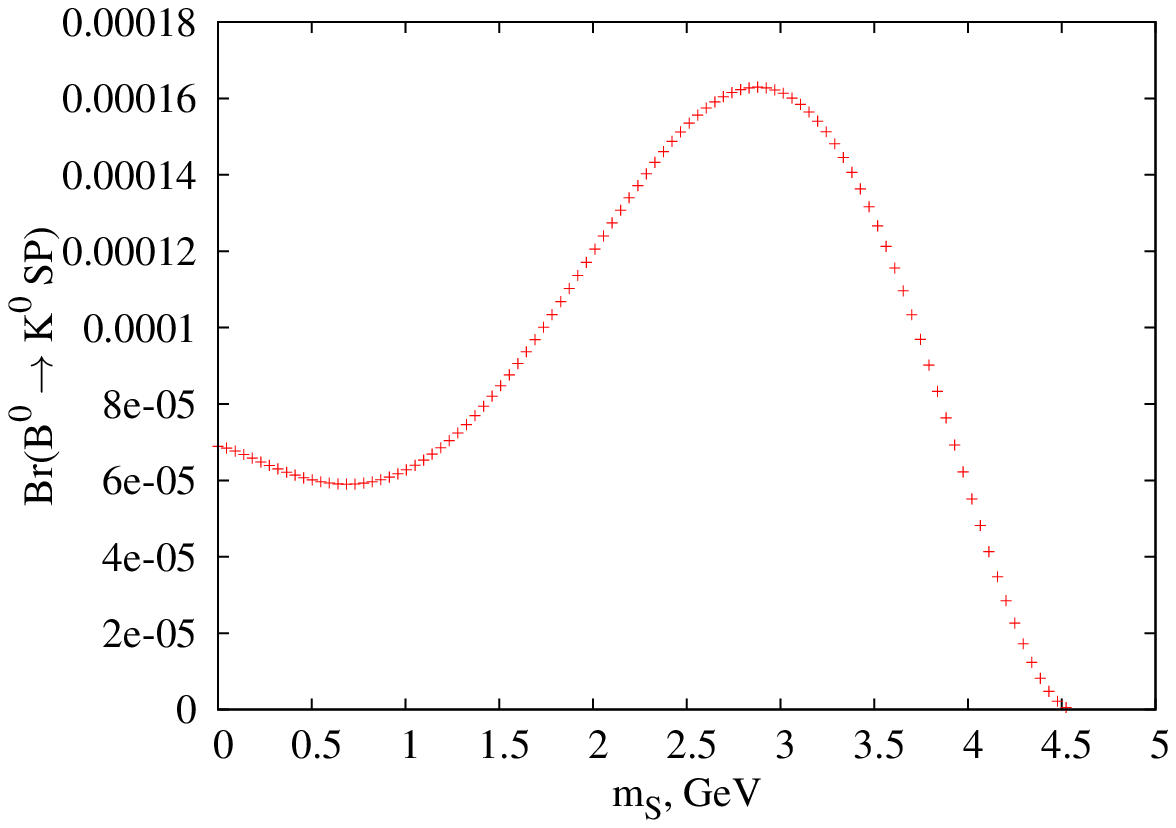} 
&
\includegraphics[width=0.5\columnwidth]{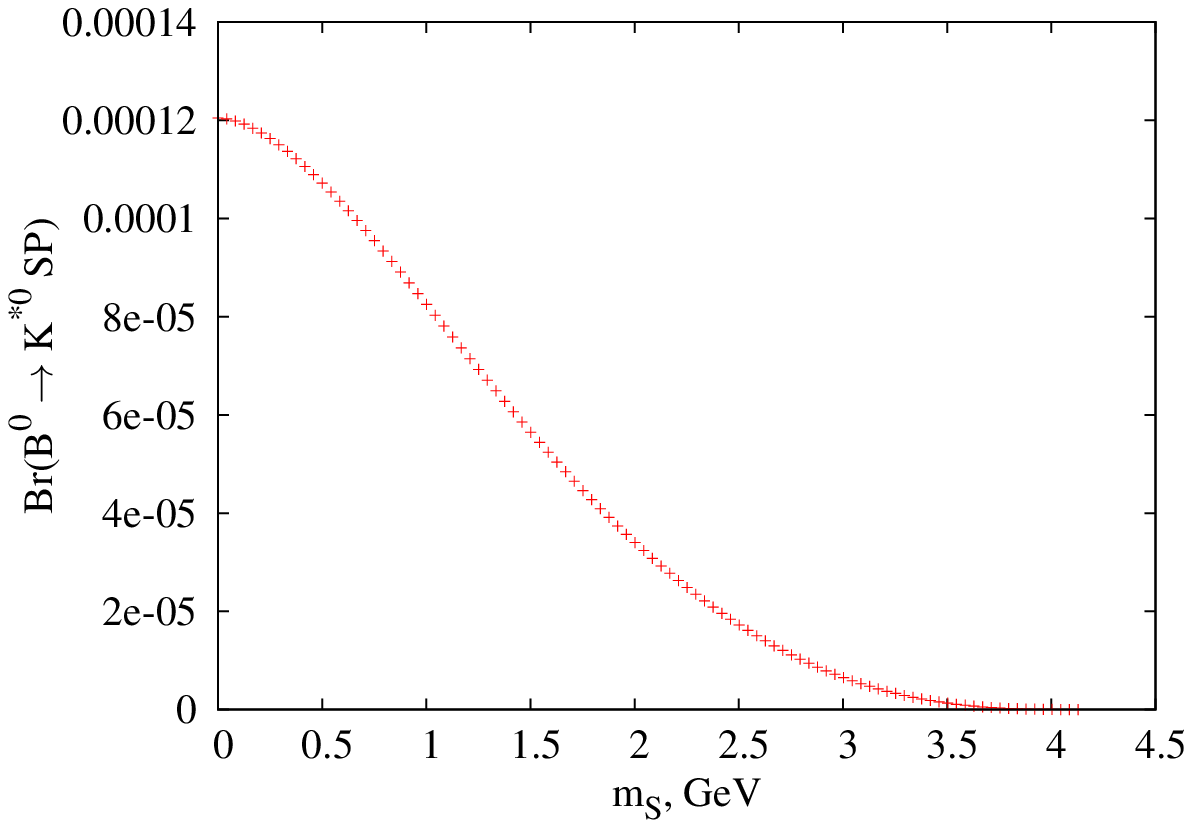} 
\end{tabular}
\end{center}
\caption{\label{taus}
Branching ratios of $B^0\to K^0SP$ (left) and
$B^0\to K^{*0}SP$ (right) as functions of scalar sgoldstino mass
$m_S$. Other parameters are as in Fig.~\ref{SP}.
}
\end{figure}

As an example of pseudoscalar meson in the final state,  
we consider decay $B^0\to K^0SP$ which is driven by the same flavor
violating squark soft terms which trigger $B_s\to SP$. We use the
following notations: $P, p_1, p_2, p_3$ are four-momentums of $B^0$,
$K^0$, $S$ and $P$; $m_{ij}^2=\l p_{i}+p_{j}\r^2$ are the Dalitz
variables. The relevant hadronic matrix element is parametrized by two 
form factors (see Appendix \ref{appendix-form-factors})
as~\cite{hep-ph/0310359} 
\[
\langle K^0|\bar{s}\gamma^{\mu}b|B^0\rangle =
\l P+p_{1}\r^{\mu}f_{+}\!\l m_{23}^2\r + \l P-p_1\r ^{\mu}f_{-}\!\l
m_{23}^{2}\r. 
\]
Then the squared matrix element of meson decay reads
\be
\label{3-body}
\begin{split}
&|{\cal M}|^2 = 
\frac{\left|\tilde{m}^{LL^{2}}_{D_{23}} -
  \tilde{m}^{RR^{2}}_{D_{23}}\right|^2}{16F^4}
\\
&\times\left[\l m_{12}^2 \!-\! m_{13}^2\r f_{+}\!\l m_{23}^2\r +
  \l m_{2}^2 \!-\! m_{3}^2\r f_{-}\!\l m_{23}^2\r \right]^2. 
\end{split}
\ee
The decay rate is obtained by integrating $|{\cal M}|^2$ over the
phase space as explained in Appendix \ref{appendix-phase-space}.  
For numerical estimates we use the set of parameter values chosen in
Sec.\,\ref{Sec:Lagrangian}; 
the decay branching ratio for $m_P=214$\,MeV as a function of $m_S$ is
presented on left panel in Fig.\ref{taus}. 

Similarly to the two-body decays studied in Sec.\,\ref{Sec:2-body} the
obtained branching ratio grows (but rather slowly) with $m_S$ in
intermediate range of masses, because
the matrix element contains a term with similar behavior, see
Eq.\,\eqref{3-body}.


Let us proceed with an example of heavy meson three-body decay into
sgoldstino pair and vector meson, ${\cal P}\to {\cal V}SP$.      
We consider decay $B^0\to K^{\star 0}SP$ which goes due
to the same sgoldstino flavor violating couplings as $B_s\to SP$. We 
adopt the following notations: $P, p_1, p_2, p_3$ are four-momentums of
$B^0$, $K^{\star 0}$, $S$ and $P$; $\epsilon$ is polarization 4-vector
of outgoing meson, $m_{ij}^2=\l p_{i}+p_{j}\r^2$ are the Dalitz
variables. We use the following hadronic matrix elements~\cite{hep-ph/0310359} 
\begin{align*}
&\langle K^*|\bar{s}\gamma^{\mu}b| B\rangle
=
\epsilon^\mu_{\;\;\nu\alpha\beta}\,
\epsilon^{*\nu}\l P+p_1\r^{\alpha}\l p_2+p_3\r^{\beta}\,g\!\l m_{23}^2\r,
\\
&\langle K^*|\bar{s}\gamma^{\mu}\gamma^{5}b|B\rangle
= -i\left\{\epsilon^{*\mu}\,f\!\l m_{23}^2\r 
+\l P+p_1\r \epsilon^* \right.
\\
&\left.
\times
\left[\l P+p_1\r_{\mu}\,a_{+}\!\l q_{23}^2\r +
  \l p_2+p_3\r_{\mu}\,a_{-}\!\l q_{23}^2\r\right]\right\},
\end{align*}
with form factors described in Appendix\,\ref{appendix-form-factors}. 
The decay amplitude can be written as follows
\begin{align*}
&{\cal M}(B\to K^*SP) = -i\beta_1(p_2-p_3)^{\mu}\langle
K^*|\bar{s}\gamma^{\mu}\gamma^{5}b|B\rangle 
\\
&+\beta_2(p_2-p_3)^{\mu}\langle K^*|\bar{s}\gamma_{\mu}b|
B\rangle\;,
\end{align*}
where $\beta_1 = \frac{\tilde{m}^{LL^{2}}_{D_{23}} +
  \tilde{m}^{RR^{2}}_{D_{23}}}{4F^2} $ and 
$\beta_2 = \frac{\tilde{m}^{LL^{2}}_{D_{23}} -
  \tilde{m}^{RR^{2}}_{D_{23}}}{4F^2} $. We present the formulas below
for the case of real $\alpha$ and $\beta$, which corresponds to
P-conservation in squark soft terms. For the squared matrix element of 
three-body decay one obtains  
\begin{widetext}
\begin{align*}
\left|{\cal M}\right|^2 = |\beta_1|^2\left\{\left[ m_{23}^2 - 2m_{2}^2 - 2m_{3}^2 +
\frac{1}{4m_1^2}\l m_{12}^2-m_{13}^2-m_2^2+m_3^2\r^2\right] f^2\!\l m_{23}^2\r
+2f\!\l m_{23}^2\r 
\right.
\\
\left.
\times\left[ m_3^2-m_2^2+\frac{1}{4m_1^2}\l M^2-m_1^2-m_{23}^2\r \l 
m_{12}^2-m_{13}^2-m_2^2+m_3^2\r\right] 
\left[\l m_{12}^2-m_{13}^2\r a_{+}\!\l m_{23}^2\r +
\l m_2^2-m_3^2\r a_{-}\!\l m_{23}^2\r\right] 
\right.\\
\left.
+
\left[ 
\frac{1}{4m_1^2}\l M^2-m_1^2-m_{23}^2\r^2 -M^2\right]
\left[\l m_{12}^2-m_{13}^2\r a_{+}\!\l m_{23}^2\r +
\l m_2^2-m_3^2\r a_{-}\!\l m_{23}^2\r\right]^2
\right\}
\\
+4|\beta_2|^2\left\{4m_1^2m_2^2m_3^2 - m_3^2 \l m_{12}^2-m_1^2-m_2^2\r^2
-m_1^2\l m_{23}^2-m_2^2-m_3^2\r^2 - m_2^2\l m_{13}^2-m_{1}^2-m_3^2\r^2
\right.
\\
\left.
+\l m_{12}^2-m_1^2-m_2^2\r \l m_{23}^2-m_2^2-m_3^2\r \l m_{13}^2-m_{1}^2-m_3^2\r
\right\}g^2\!\l m_{23}^2\r\;,
\end{align*}
\end{widetext}
where $M$ stands for the mass of decaying meson. 
The decay rate is obtained by integrating of $|{\cal M}|^2$ over the
phase space as explained in Appendix \ref{appendix-phase-space}.  
For numerical estimates we use the set of parameter values introduced
in Sec.\,\ref{Sec:Lagrangian};
the decay branching ratio for $m_P=214$\,MeV is depicted on right
panel of Fig.\ref{taus} 
for kinematically allowed values of scalar
sgoldstino mass $m_S$. 

Several comments are in order. 
The decay rates for charged mesons, e.g. $B^+\to K^+SP$, are given by
the same formulas as presented above. The decay rates for $\bar{B}^0$
(and $B^-$) meson are given by the same formulas with replacement
$\tilde{m}^{XX^2}_{D_{23}}\to \tilde{m}^{XX^2}_{D_{32}}$. 
Likewise it is 
straightforward to write down formulas for vector meson decays into
sgoldstino and for various three-body decays of pseudoscalar and
vector $D$-mesons and $B_c$ mesons.

\section{$t\to c(u) SP$ and $\tau\to\mu(e) SP$}

Flavor-violating couplings of sgoldstino to top-quark in
Eq.\,\eqref{flavor-violation} result in decays $t\to c SP$ and $t \to
u  SP$. To describe it below we use the following notations: 
$P, p_1, p_2, p_3$ are four-momentums of $t$-quark, lighter up-type
quark ($c$ or $u$), $S$ and $P$; $M$ denotes mass of top-quark,
$m_{ij}^2=\l p_{i}+p_{j} \r^2$ are the Dalitz variables. The squared
matrix element of the three-body decay can be written as 
\be
\label{top_ampl}
\left|{\cal M}\right|^2 = 3\cdot\left[\l |\gamma_1|^2+|\gamma_2|^2\r \tau_1 
+ \l|\gamma_2|^2-|\gamma_1|^2\r\tau_2\right]\,,
\ee
where overall numerical factor 3 refers to the number of colors, 
\begin{align*}
\tau_1 &= \l m_{12}^2-m_{13}^2\r^2-\l m_2^2-m_3^2\r ^2
\\&-\l 2m_2^2+2m_3^2-m_{23}^2\r\l M^2 + m_1^2-m_{23}^2\r\;,
\\
\tau_2 & = 2Mm_1\l 2m_2^2+2m_3^2-m_{23}^2\r\;,
\end{align*}
and $\gamma_1 =
\frac{\tilde{m}^{LL^{2}}_{U_{23}}+\tilde{m}^{RR^{2}}_{U_{23}}}{4F^2}$, 
$\gamma_2 =
\frac{\tilde{m}^{LL^{2}}_{U_{23}}-\tilde{m}^{RR^{2}}_{U_{23}}}{4F^2}$ for
$t\to c SP$ and $\gamma_1 = 
\frac{\tilde{m}^{LL^{2}}_{U_{13}}+\tilde{m}^{RR^{2}}_{U_{13}}}{4F^2}$,
$\gamma_2 =
\frac{\tilde{m}^{LL^{2}}_{U_{13}}-\tilde{m}^{RR^{2}}_{U_{13}}}{4F^2}$
for $t\to u SP$. 

\begin{figure}[htb]
\begin{center}
\begin{tabular}{ll}
\includegraphics[width=0.5\columnwidth]{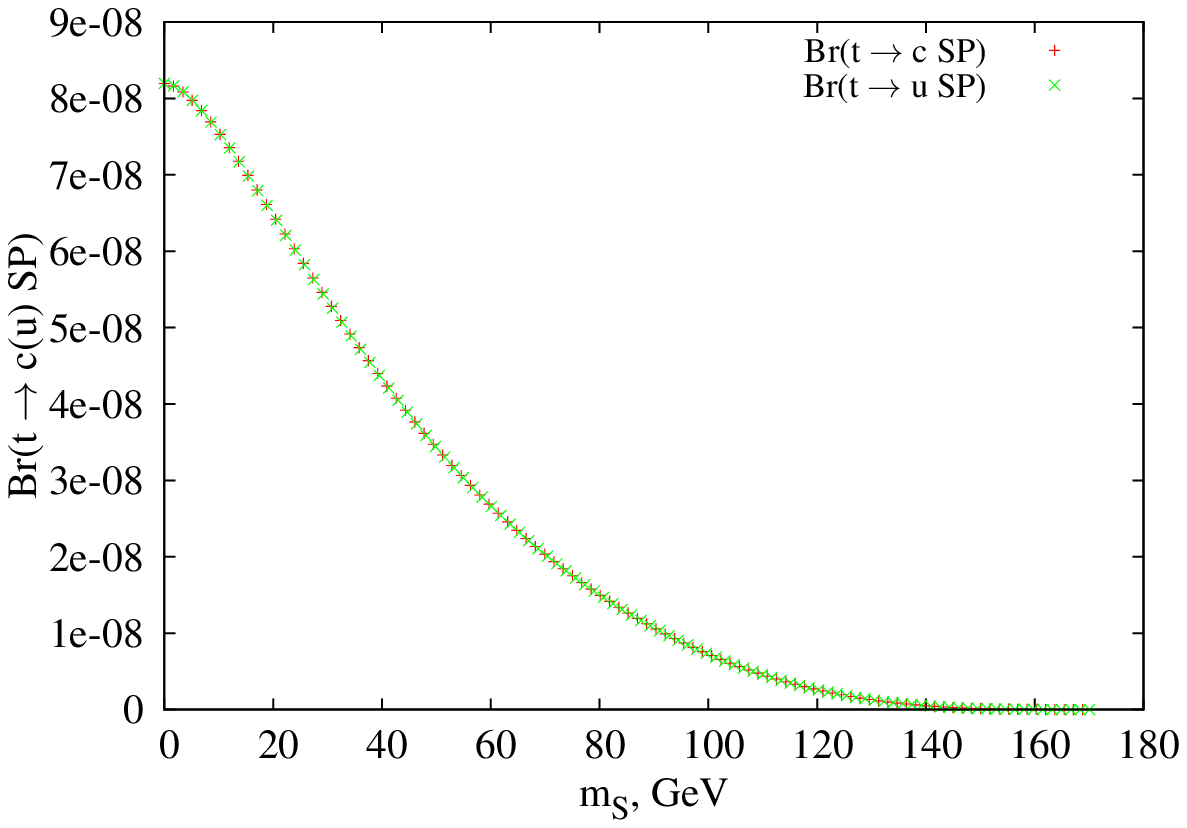} 
&
\includegraphics[width=0.5\columnwidth]{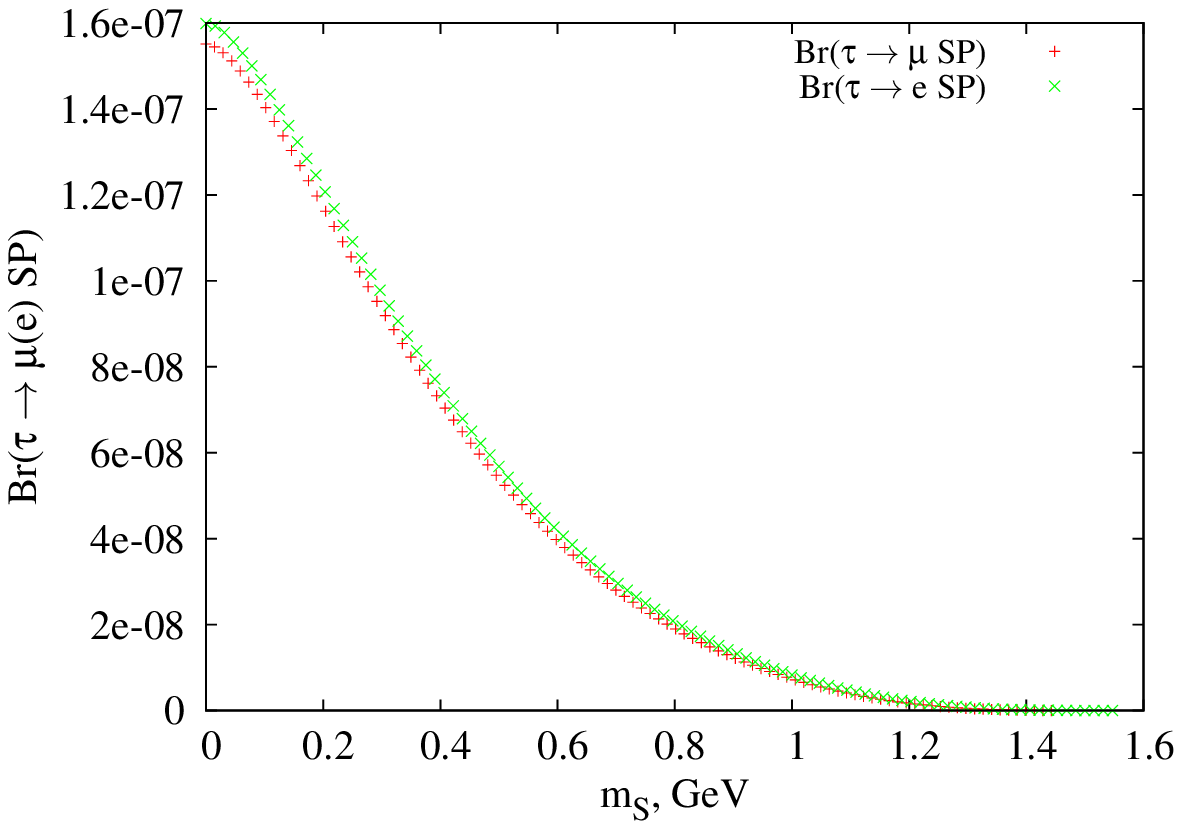} 
\end{tabular}
\end{center}
\caption{\label{top}
Branching ratios for $t \to c(u) SP$ (left) and 
$\tau \to \mu(e) SP$ (right) as functions of $m_S$. Other parameters
are as in Fig.~\ref{SP}.
}
\end{figure}

For numerical estimates we use the set of values chosen in
Sec.\,\ref{Sec:Lagrangian}. For the mass and width of top-quark we
take 172.0\,GeV and 1.3\,GeV respectively. Integrating  $|{\cal M}|^2$
over the phase space as explained in
Appendix~\ref{appendix-phase-space} one obtains the decay branching 
ratios which are presented  in Fig.~\ref{top} (left) as functions of
$m_S$ for $m_P=214$\,MeV. 


Decays $\tau\to\mu SP$ and $\tau\to e  SP$ can be treated in the
similar way. Now the notations are as follows: $P, p_1, p_2, p_3$
are four-momentums of $\tau$-lepton, light lepton $l$ (muon $\mu$ or
electron $e$), $S$ and
$P$; $M$ stands for $\tau$-lepton mass, $m_{ij}^2=\l p_{i}+p_{j}\r^2$
are the Dalitz variables. The squared matrix element of the three-body
decay is 
presented by Eq.~\eqref{top_ampl} without color factor 3 and with 
$\gamma_1 =
\frac{\tilde{m}^{LL^{2}}_{L_{23}}+\tilde{m}^{RR^{2}}_{L_{23}}}{4F^2}\,,\; 
\gamma_2 =
\frac{\tilde{m}^{LL^{2}}_{L_{23}}-\tilde{m}^{RR^{2}}_{L_{23}}}{4F^2} $
for decay $\tau\to\mu SP$ and 
$\gamma_1 =
\frac{\tilde{m}^{LL^{2}}_{L_{13}}+\tilde{m}^{RR^{2}}_{L_{13}}}{4F^2}\,,\; 
\gamma_2 =
\frac{\tilde{m}^{LL^{2}}_{L_{13}}-\tilde{m}^{RR^{2}}_{L_{13}}}{4F^2}$
for decay $\tau\to eSP$.
The decay rates are obtained by integrating $|{\cal M}|^2$ over the
phase space as explained in Appendix \ref{appendix-phase-space}.  
For numerical estimates we adopt the parameter values 
given in Sec.\,\ref{Sec:Lagrangian};
the decay branching ratios for $m_P=214$\,MeV are presented in
Fig.~\ref{top} (right)
as functions of $m_S$.

\section{Summary}
\label{Sec:Summary}

To conclude we have studied sgoldstino pair production by heavy quark
and $\tau$-lepton decays. The obtained estimates of corresponding branching
ratios for $B$- and $D$-mesons, $t$-quark and $\tau$-lepton decays show that
the processes are available for study already with collected statistics
in Belle, BaBar, LHCb, CLEOc. The numerical results are presented for
a particular set of supersymmetry parameters, see Sec.
\ref{Sec:Lagrangian}. These are mixing angles in squark (slepton)
squared mass matrices $\delta$, superpartner mass scale $\tilde{m}$
and supersymmetry breaking parameter $F$.  All the branching ratios
discussed above scale as $\propto \delta^2 \cdot \tilde{m}^4/F^4$,
which allows to give predictions for other phenomenologically
  viable values of the model parameters. 

Note that generally, soft supersymmetry breaking terms violates
CP-symmetry, that, in particular, splits rates of sgoldstino pair
production in particle (meson) and antiparticle (antimeson) decays,
e.g. 
\[
\Br\l B_s\to SP\r\neq\Br\l \bar{B}_s\to SP\r\;.
\] 
Similar statement holds for the three-body decays. In case of two-body
decays the splitting (so-called asymmetries) can be estimated with the
help of decay {\em rate} formulas presented in
Sec.\,\ref{Sec:2-body}. Three-body decay rates with CP-violation
can be calculated with help of formulas for {\em amplitudes} given in 
Sec.\,\ref{Sec:3-body}.

{\bf Acknowledgments}

We thank A. Golutvin for stimulating questions and N. Nikitin for
discussions. This work is partially supported by the grants of the 
President of the Russian Federation NS-5590.2012.2, MK-2757.2012.2
(S.D.) and by Russian Foundation for Basic Research 11-02-01528-a
(S.D. and D.G.) and 11-02-92108-YAF\_a (D.G.). The work of D.G. is
supported in part by SCOPES. 

\appendix
\section{Meson Form factors used in numerical estimates}
\label{appendix-form-factors}
For meson leptonic decay constants we 
use $f_{D}=207$~MeV~\cite{arXiv:0904.1895},
$f_{B}=190$~MeV~\cite{arXiv:0902.1815}, 
$f_{B_s}=231$~MeV~\cite{arXiv:0902.1815}. For other meson form factors 
we adopt the universal form 
of dependence on momentum transfer $q^2$ \cite{hep-ph/0310359}
\[
F\l q^2\r = \frac{F\l 0\r}{1 - a\l q^2/M^2\r + b\l q^2/M^2\r^2}
\]
with values of dimensionless parameters listed in the 
Table~\ref{formfactors}. 
\begin{table}[!htb]
\begin{center}
\begin{tabular}{|c|c|c|c|c|c|c|}
\hline
$F$  & $F_{1}^{BK}$ & $F_{0}^{BK}$ & $V^{BK}$ & $A_0^{BK^*}$ &
$A_{1}^{BK^*}$ & $A_{2}^{BK^*}$  \\\hline
$F(0)$  & 0.35 & 0.35 & 0.31 & 0.31 & 0.26 & 0.24 \\
$a$    & 1.58 & 0.71 & 1.79 & 1.68 & 0.93 & 1.63 \\
$b$    & 0.68 & 0.04 & 1.18 & 1.08 & 0.19 & 0.98 \\\hline
\end{tabular}
\caption{Numerical coefficients 
in meson formfactors~\cite{hep-ph/0310359}.
\label{formfactors} 
}
\end{center}
\end{table}
These formfactors are related to those entering formulas 
in the main text, as follows~\cite{hep-ph/0310359}
\begin{eqnarray*}
F_1\!\l q_{23}^2\r = f_{+}\!\l q_{23}^2\r,\;
F_0\!\l q_{23}^2\r = \frac{q_{23}^2}{ M^2-m_1^2}\,f_{-}\!\l
q_{23}^2\r ,\\
V \!\l q_{23}^2\r = -\l M+m_1\r g\!\l q_{23}^2\r,\;\;
A_1\!\l q_{23}^2\r = -\frac{f\l q_{23}^2\r}{M+m_1},\\
A_2\!\l q_{23}^2\r = \l M+m_1\r a_{+}\!\l q_{23}^2\r, \\
A_3\!\l q_{23}^2\r - A_{0}\!\l q_{23}^2\r  =
\frac{q_{23}^2}{2m_{1}}a_{-}\!\l q_{23}^2\r,
\end{eqnarray*}
where
\[
A_{3}\l q_{23}^2\r = \frac{M+m_1}{2m_{1}}\,A_{1}\!\l q_{23}^2\r -
\frac{M-m_1}{2m_1}\,A_2\!\l q_{23}^2\r\;.
\]

\section{Three-body phase space integration}
\label{appendix-phase-space}

To calculate the rates $\Gamma$ of 
the three-body decays described in the main text 
one has to integrate presented there 
squared matrix elements $|{\cal
  M}|^2$ over the three-body phase space of corresponding final state.
For the decay of particle of
mass $M$ into three particles of masses $m_i$ with 4-momenta $p_i$, 
$i=1,2,3$, one has the general formula (see e.g. Sec.\,39 in
Ref.\cite{FERMILAB-PUB-10-665-PPD})  
in terms of the Dalitz
variables $m_{ij}^2=\l p_i+p_j\r^2$, 
\begin{equation*}
\Gamma = \frac{1}{(2\pi)^3}\frac{1}{32M^3} 
\int_{(m_{12}^2)_{min}}^{(m_{12}^2)_{max}}
\!\!dm_{12}^2
\int_{(m_{23}^2)_{min}}^{(m_{23}^2)_{max}}\!\!dm_{23}^2|{\cal M}|^2,
\end{equation*}
where $(m_{12}^2)_{min} = (m_1+m_2)^2$, $(m_{12}^2)_{max} =
(M-m_3)^2$,  
\begin{align*}
(m_{23}^2)_{min}& = (E_{2}^{*}\!+\!E_{3}^{*})^2 -
\left(\!\!\sqrt{E_{2}^{*2}\!-\!m_{2}^2} + \sqrt{E_{3}^{*2}\!-\!m_{3}^2}\right)^{\!2}\!,
\\
(m_{23}^2)_{max}& = (E_{2}^{*}\!+\!E_{3}^{*})^2 -
\left(\!\!\sqrt{E_{2}^{*2}\!-\!m_{2}^2} - \sqrt{E_{3}^{*2}\!-\!m_{3}^2}\right)^{\!2}\!,
\end{align*}
and $E_{2}^{*} = \frac{m_{12}^2 - m_{1}^2 + m_{2}^2}{2m_{12}}$, 
$E_{3}^{*} = \frac{M^2-m_{12}^2-m_{3}^2}{2m_{12}}$ .


\end{document}